\documentclass[conference]{IEEEtran}
\IEEEoverridecommandlockouts
\usepackage{amsmath,amssymb,amsfonts}
\usepackage{algorithmic}
\usepackage{graphicx}
\usepackage{textcomp}
\usepackage[dvipsnames]{xcolor}
\def\BibTeX{{\rm B\kern-.05em{\sc i\kern-.025em b}\kern-.08em
    T\kern-.1667em\lower.7ex\hbox{E}\kern-.125emX}}

\usepackage[style=ieee, maxbibnames=999, maxcitenames=2, mincitenames=1]{biblatex}
\usepackage[hidelinks]{hyperref}
\usepackage{url}
\usepackage{cleveref}
\usepackage{booktabs}
\usepackage{multirow}
\usepackage{comment}
\usepackage{mathtools, cuted}
\usepackage{xspace}
\usepackage{makecell}
\usepackage{soul}
\usepackage{flushend}

\usepackage{enumitem}

\usepackage{array}
\newcolumntype{P}[1]{>{\centering\arraybackslash}p{#1}}
\newcolumntype{M}[1]{>{\centering\arraybackslash}m{#1}}

\newcommand{\ttbs}{TTBS\xspace}

\newcommand{\base}{\textit{baseline}\xspace}
\newcommand{\power}{\textit{power-aware}\xspace}

\newcommand{\SoBigDataITAck}{European Union -- NextGenerationEU -- National Recovery and Resilience Plan (Piano Nazionale di Ripresa e Resilienza, PNRR) -- Project: ``SoBigData.it -- Strengthening the Italian RI for Social Mining and Big Data Analytics'' -- Prot. IR0000013 -- Avviso n. 3264 del 28/12/2021\xspace}

\newcommand{\rechargeAck}{Italian Government (Ministero dell'Università e della Ricerca, PRIN 2022 PNRR) -- Project: ''RECHARGE: monitoRing, tEsting, and CHaracterization of performAnce Regressions`` -- Decreto Direttoriale n. 1205 del 28/7/2023\xspace}

\def\eg{\emph{e.g.},\xspace} 

\def\ie{\emph{i.e.},\xspace}

\makeatletter
\let\MYcaption\@makecaption
\makeatother

\usepackage[font=footnotesize]{subcaption}

\makeatletter
\let\@makecaption\MYcaption
\makeatother

\begin{document}

\title{Exploring sustainable alternatives for the deployment of microservices architectures in the cloud}

\author{\IEEEauthorblockN{Vittorio Cortellessa}
\IEEEauthorblockA{\textit{SPENCER Lab} \\
\textit{University of L'Aquila, Italy}\\
vittorio.cortellessa@univaq.it}
\and
\IEEEauthorblockN{Daniele {Di Pompeo}}
\IEEEauthorblockA{\textit{SPENCER Lab} \\
\textit{University of L'Aquila, Italy}\\
daniele.dipompeo@univaq.it}
\and
\IEEEauthorblockN{Michele Tucci}
\IEEEauthorblockA{\textit{SPENCER Lab} \\
\textit{University of L'Aquila, Italy}\\
michele.tucci@univaq.it}
}

\maketitle

\begin{abstract}
As organizations increasingly migrate their applications to the cloud, the optimization of microservices architectures becomes imperative for achieving sustainability goals. Nonetheless, sustainable deployments may increase costs and deteriorate performance, thus the identification of optimal tradeoffs among these conflicting requirements is a key objective not easy to achieve. This paper introduces a novel approach to support cloud deployment of microservices architectures by targeting optimal combinations of application performance, deployment costs, and power consumption. By leveraging genetic algorithms, specifically NSGA-II, we automate the generation of alternative architectural deployments.
The results demonstrate the potential of our approach through a comprehensive assessment of the Train Ticket case study.

\end{abstract}


\section{Introduction}\label{sec:intro}

As the world's reliance on digital technologies grows, so does the environmental impact of data centers.
The escalating demands for computing resources, coupled with the energy-intensive operations of data centers, significantly contribute to their substantial carbon footprint~\cite{BELKHIR2018448}. 
Cloud computing is at the center of these demands, as it has emerged as a key solution to enhance technological capabilities of companies.
The prevailing trend in cloud deployment involves designing applications based on the principles of microservices architecture~\cite{Rodriguez-Gracia2019Microservices}.
This architectural approach is widely favored for cloud environments, as it aligns seamlessly with the dynamic and scalable nature of cloud computing~\cite{10.1145/3183440.3194991}.

When deploying microservices applications in the cloud, the energy footprint of the application is frequently overlooked, whereas prioritizing concerns such as performance and deployment costs, which are more closely related to operational success~\cite{Mytton2020Assessing}.
The challenge of considering energy consumption stems from the introduction of an additional dimension into deployment planning, and from the inherent complexity in evaluating the energy requirements across various potential alternative configurations~\cite{HOUSSEIN2021100841,10.1145/2788397,spe.2409}.
Nonetheless, nowadays it is essential to consider sustainability as a trade-off of performance, deployment cost, and energy consumption \cite{Huppes2005A} to ensure the long-term viability of cloud-based microservices architectures.

A number of approaches \cite{Xu2016EnReal:,Baliga2011Green,Armstrong2017Towards} emerged in recent years to optimize energy consumption and cost when deploying to the cloud.
Nevertheless, these approaches seldom address this problem at the architectural level and, as a consequence, often they lack the capability to empower designers with a comprehensive understanding of the intricate trade-offs emerging from this task.
This lack of understanding is further exacerbated by the fact that the energy consumption of a microservices architecture is not only a function of the deployment configuration but also of the user behavior~\cite{Hahnel2017Application}.

In this paper, we aim to address this lack by presenting a novel approach to explore sustainable solutions when deploying microservices architectures in the cloud.
Specifically, we exploit NSGA-II~\cite{DBLP:conf/ppsn/DebAPM00} to generate diverse deployment configurations through the application of refactoring actions to an initial architecture.
Our method provides the designer with design alternatives that represent optimal trade-offs of system performance, deployment costs, and power consumption.
Moreover, we support the exploration of the intricate relationship between power consumption, cost and user behavior by observing the distribution of these factors on different types of user requests.
Finally, we investigate how architectural solutions change when optimizing for power consumption, thus aiming to discern recurring refactoring actions in a power-aware context.

To enhance the practical significance of our research, we showcase our approach on the Train Ticket Booking Service case study~\cite{10.1145/3183440.3194991}, which is a widely used microservices benchmark designed to reflect a real-world scenario.

Our results reveal that the introduction of a power consumption objective significantly affects system response time and, contrary to conventional expectations, has a negligible impact on deployment costs.
Also, the contribution of individual types of requests greatly varies depending on the deployment decisions. In perspective, our approach can identify opportunities for merging smaller microservices to save on power consumption and costs.

The remainder of this paper is organized as follows.
\Cref{sec:approach} presents the proposed approach.
\Cref{sec:experimental_setup} describes our settings and experiments.
\Cref{sec:rq1,sec:rq2,sec:rq3} discuss the research questions and present the results.
\Cref{sec:related} presents the related work.
\Cref{sec:t2v} presents the threats to validity.
\Cref{sec:conclusion} concludes the paper and discusses future work.
 \section{Approach}\label{sec:approach}

Our approach is based on genetic algorithms, which are bio-inspired algorithms that  achieve the desired near-optimal Pareto fronts by evolving the specie through crossover and  mutation operators.
The specie in our context is the set of candidate architectural solutions generated through refactoring, and the genetic algorithm is NSGA-II~\cite{DBLP:conf/ppsn/DebAPM00}.
NSGA-II is widely used in the literature and it is considered one of the most effective multi-objective evolutionary algorithms~\cite{DBLP:journals/mta/KatochCK21}.
Its main characteristic is the use of a fast non-dominated sorting approach to select the best individuals in the population.
As a result, NSGA-II is able to achieve a good convergence and diversity of the Pareto front.

An individual in our population is a sequence of refactoring actions.
Each action is applied to an element in the software architecture.
When combining refactoring actions to compose the sequence, the algorithm must respect constraints (i.e., pre- and post-conditions) that are specific to each action.
Therefore, once an individual has been created, it can be applied to the architecture to generate a new architectural alternative.

\subsection{Objectives}\label{sec:approach:objective_computation}

The goal of this approach is to generate alternative architectures with lower power consumption, while preserving, or even improving, the performance, by minimizing the overall economic cost of deployment and the refactoring effort to change the architecture. 
In the following, we describe the objectives that are considered in this study.

\paragraph{Power consumption}

\textcite{Xu2023} proposed a model to estimate the total power consumption of a server in case its power consumption is only known when it is fully utilized.
By considering that most of the power consumption of a server comes from its CPU~\cite{BELOGLAZOV2012755,10.1145/1250662.1250665,10.1145/1273440.1250665}, \citeauthor{Xu2023} in their model compute the power consumption of a server as the combination of power consumption of the CPU when used and when idle, where the latter one is obtained by scaling down the former by a factor.

We adopt their model, thus we estimate the power consumption of an application as the sum of the power consumption of all the nodes (i.e., servers on which it is deployed), as follows:
\[ power = \sum_{n \in N}(1-U_n) \cdot k \cdot power_{max}(n) + U_n \cdot power_{max}(n)  \]
where $n$ is a node among the $N$ used ones, $U_n$ is its utilization, $power_{max}(n)$ is the maximum power consumed when $n$ is fully utilized, $k$ is the scaling factor, introduced by \citeauthor{Xu2023}, by which $power_{max}$ is reduced when the server is idle.

\paragraph{Response Time}

Performance of a software system is a broad term that can be quantified by several metrics. In this study, we consider system response time as the performance metric to be minimized. It can be obtained as the combination of response times of the system when it is triggered by different types of requests. This metric is a common outcome of tools that solve performance models.

\paragraph{Complexity}

Refactoring actions can be more or less complex to be applied on a software architecture. In this study, we introduce a complexity metric defined as follows:
\[ complexity = \sum_{a=1}^{l} C_{base}(a) \cdot C_{arch}(a,e) \]
where $C_{base}(a)$ is the base complexity associated to a type of action $a$, and $C_{arch}(a,e)$ is the architectural complexity that is computed on the basis of the element $e$ on which the refactoring is applied, as explained in the following.
$C_{base}$ can be estimated through several estimation models~\cite{boehm2009software,trendowicz2013software}.
Often, these models exploit knowledge of the software system and implement business rules that are hard to generalize.
For this reason, we estimated the base complexity of each type of action on the basis of prior studies on software architecture refactoring~\cite{DBLP:conf/euromicro/Pompeo022,DBLP:conf/sbcars/RagoVDFH17}.
The $C_{arch}(a,e)$ complexity, instead, is related to the target element of the refactoring action, because a type of action can have different complexities when applied to different target elements of the architecture.
For example, the more interconnected is a target element, the higher is the architectural complexity of the refactoring action.
Or, in other words, target elements that are used relatively less than other ones induce a lower complexity, due to the lighter impact that they have on the rest of the architecture.

\paragraph{Cost}

The deployment cost of a software architecture in a cloud based infrastructure is the cost of the hardware nodes on which it is deployed.
Thus, it is related to the number and the type of nodes used to deploy the architecture.

Hence, we considered the overall cost of the software architecture as the sum of the hourly cost of all the used nodes ($N$), as follows:
\[ cost = \sum_{n \in N} cost(n) \]

\subsection{Modeling assumptions}\label{sec:modeling-assumption}

In this study, we use UML to build a software architecture of a system.
In particular, we use a Component Diagram to model static connections among software components, Sequence Diagrams to model the dynamic behavior of the software system, and Deployment Diagram to model the hardware platform and the component allocation.
\begin{figure}[htbp]
    \centering
    \begin{subfigure}[b]{0.26\columnwidth}
        \includegraphics[width=\textwidth]{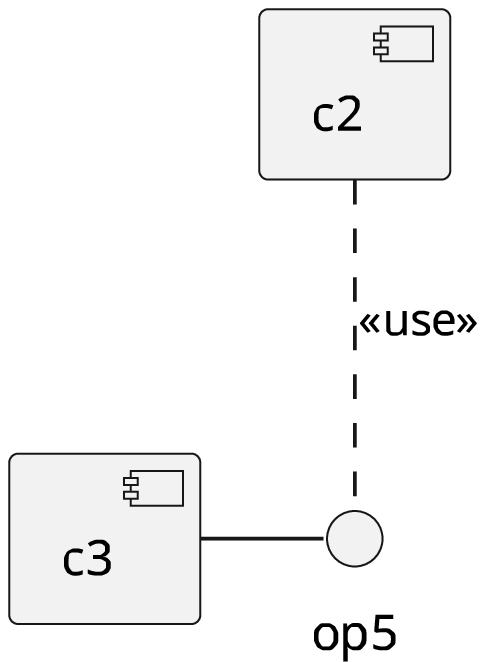}
        \caption{Component}
        \label{fig:uml-component-diagram}
    \end{subfigure}
    \hfill
    \begin{subfigure}[b]{0.4\columnwidth}
        \includegraphics[width=\textwidth]{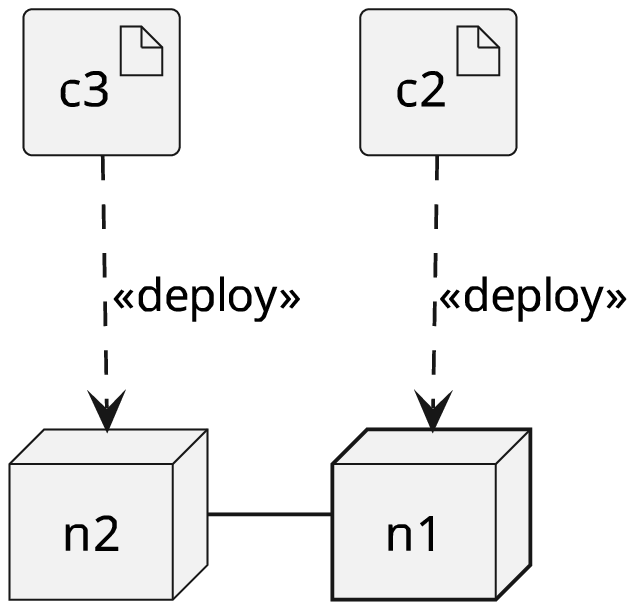}
        \caption{Deployment}
        \label{fig:uml-deployment-diagram}
    \end{subfigure}
    \hfill
    \begin{subfigure}[b]{0.26\columnwidth}
        \includegraphics[width=\textwidth,trim={0 3cm 0 0},clip]{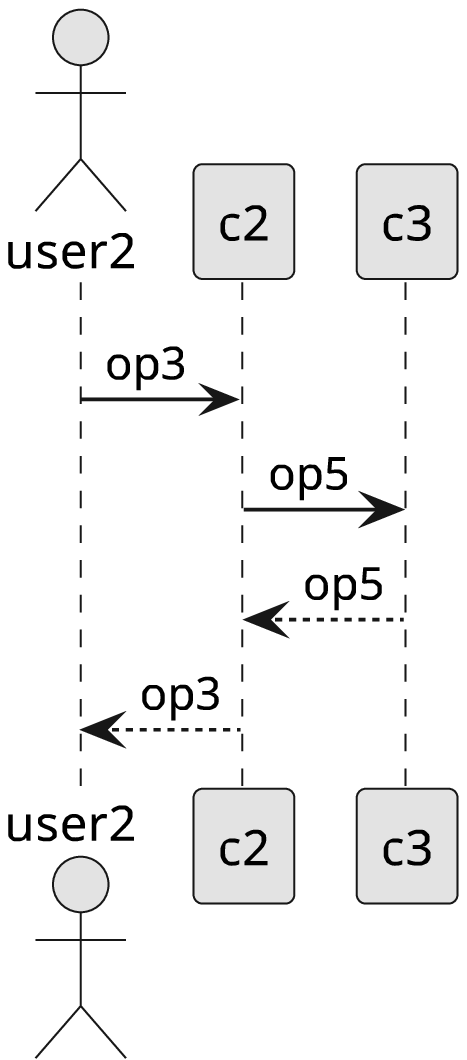}
        \caption{Sequence}
        \label{fig:uml-sequence-diagram}
    \end{subfigure}
    \hfill
    \caption{UML diagrams}
    \label{fig:uml-diagrams}
\end{figure}

\Cref{fig:uml-diagrams} shows simple examples of UML diagrams used in this paper.
Natively, UML does not provide notation elements to model performance, but the MARTE profile \cite{MARTE} has been introduced for this goal.
In this study, we use Layered Queueing Networks (LQN)~\cite{DBLP:conf/mascots/FranksW04} to model and analyze the performance of the software architecture.
Several approaches have been introduced to transform a UML model into a performance model~\cite{DBLP:conf/qest/LiAZCP17,10.1145/3185768.3186285,DBLP:conf/qest/BortolussiGHT10}, and here we adopt the approach in~\cite{DBLP:journals/infsof/CortellessaPST23} to obtain the LQN model.

\subsection{Refactoring catalog}\label{sec:approach:refactoring_catalog}

In this study, the refactoring catalog is made up of refactoring actions that were conceived to improve software performance.
We use the five types of refactoring actions described in the following.

\paragraph{REDO - Redeploy Existing Component}
Redeploying an existing component involves modifying the deployment by relocating a selected component to a newly created node. 
This action aims to optimize the software architecture by strategically redistributing components, while ensuring that the new node has connections with all nodes directly linked to the original one.

\paragraph{MOVE - Relocate Operation to Existing Component}
The relocation of an operation to an existing component is an action that involves selecting and transferring a specific operation to an existing target component. 
This action propagates modifications in all scenarios where the operation occurs.

\paragraph{CLON - Clone Node}
Cloning a node is an action aimed at introducing a replica of an existing node. 
It ensures that every deployed component and connection of the original node is duplicated, thus also contributing to system redundancy and fault tolerance.

\paragraph{MOTN - Move Operation to New Component on New Node}
Moving an existing operation to a new component on a new node is a complex refactoring action that requires consistency of static, dynamic and deployment views. 
This involves adding the newly created component in the dynamic view and creating a new node, artifact, and related links in the deployment view. 
Messages addressed to the moved operation are appropriately forwarded, thus ensuring the seamless integration of the new component.

\paragraph{DROP - Remove Node}
A node removing action induces the relocation of its deployed components on neighbor nodes, in order not to remotely spreading components that could be highly interconnected.

 \section{Experimental setup}\label{sec:experimental_setup}

In this section, we describe the experimental setup used to evaluate the proposed approach.

\subsection{Research questions}\label{sec:research_questions}

This study aims to answer the following research questions.

\newlist{RQlist}{enumerate}{1}

\setlist[RQlist, 1]
{label=\textbf{RQ\arabic{RQlisti}.},
leftmargin=3.2em,
rightmargin=0pt
}

\begin{RQlist}
    \item \emph{What is the impact of sustainability on performance and cost?}
    
    \underline{Rationale}: Adding sustainability constraints can negatively impact performance and cost of a software system. We aim at estimating what we need to trade for more sustainable deployments.

    \item \emph{What is the effect of refactoring actions on the distribution of power and cost across different types of requests?}

    \underline{Rationale}: The refactoring actions can change the power and cost distribution across different user behaviors (i.e., different types of requests). We observe these distributions in optimal solutions, thus providing a view that may enable architectural decisions depending on user behaviors.

    \item \emph{How do the architectural solutions change when introducing power consumption among the optimization objectives?}

    \underline{Rationale}: The introduction of power consumption among objectives may lead to use different refactoring actions. Here, we aim at studying their frequencies to identify possible recurring choices.

\end{RQlist}

\subsection{Case study}\label{sec:case_studies}

We experimented our optimization model on a software system, namely Train Ticket Booking Service (\ttbs), which is a web-based booking application whose architecture is based on the microservices paradigm~\cite{10.1145/3183440.3194991}. 
The system is made up of 40 microservices, each one deployed on a Docker container. 
Among all the \ttbs scenarios that a user can trigger, we focused on the most common and critical ones.
In particular, we consider the following scenarios: \emph{Login}, \emph{Update user details}, and \emph{Rebook a ticket}.

The architectural specification, including component, deployment, and sequence diagrams (in UML format), and the analytical models (LQN) of \ttbs are provided for experiment replication\footnote{Dataset: \url{https://zenodo.org/doi/10.5281/zenodo.10246197}}.

\subsection{Experimental design}\label{sec:experimental_design}
\begin{table*}[hbtp]
    \centering    
    \caption{Instance Types and Descriptions. Each Amazon EC2 instance type has been labeled with respect to \emph{Speed Factor}, \emph{Power Consumption} in Watt, and \emph{Cost} in USD/h.}
    \label{tab:aws-instances}
\begin{tabular}{lm{9cm}ccc}
    \toprule
    \textbf{Instance Type} & \textbf{Description} & \textbf{Speed Factor} & \textbf{Power Consumption} & \textbf{Cost} \\
    \midrule
    \texttt{d2.2xlarge}  & Designed for resource-intensive workloads like scientific simulations, machine learning, or real-time analytics.                                                             & 4.67 & 83.4 & 0.46 \\
    \midrule
    \texttt{m6i.xlarge}  & Offers good performance for demanding web applications, database servers, or content delivery with lower energy consumption and operational cost.                            & 3.48 & 32.4 & 0.13 \\
    \midrule
    \texttt{t2.medium}   & Well-balanced for a wide range of general-purpose workloads, cost-effective for typical business applications, small to medium-sized websites, and development environments. & 2.33 & 14.1 & 0.03 \\
    \midrule
    \texttt{t2.micro}    & Best suited for non-resource-intensive tasks like simple web hosting, low-traffic blogs, or small-scale personal projects with low energy consumption and operational cost.  & 1.17 & 6.40 & 0.004 \\
    \midrule
    \texttt{m5ad.xlarge} & Suitable for tasks where a balance between energy efficiency and operational cost is desired, such as basic web hosting or non-resource-intensive background tasks.          & 1.14 & 29.9 & 0.25 \\
    \bottomrule
    \end{tabular}
\end{table*}
\paragraph{Setting the model} First, we built a UML model of the \ttbs.
Then, we augmented the model with the MARTE profile~\cite{MARTE}.
In particular, we used the \emph{GQAM}, \emph{HwLayout}, and \emph{GRM} packages to specify the performance, power, and cost properties, respectively.

With regard to performance properties, we gathered data from a running \ttbs application to set performance input data (\eg operations demand vectors), as suggested in~\cite{DBLP:conf/epew/WillneckerDBSKG15,DBLP:journals/jss/CortellessaPET22}. 
Furthermore, we compute the speed factor of an instance as the PassMark average CPU mark of the used CPU\footnote{PassMark Software CPU Benchmark: \url{https://www.cpubenchmark.net/}} scaled over the number of virtual CPUs provided by the Amazon EC2 instance type.

To estimate the power consumption under full utilization of CPU (i.e., $power_{max}(n)$), we extracted data from the Amazon EC2 Instances Carbon Footprint Estimator dataset\footnote{Amazon EC2 Instances Carbon Footprint Estimator: \url{https://docs.google.com/spreadsheets/d/1DqYgQnEDLQVQm5acMAhLgHLD8xXCG9BIrk-\_Nv6jF3k/edit\#gid=224728652} [Accessed: 2023-11-28]}. From the same dataset, we have estimated the $k$ factor that we adopt to scale down the power when the CPU is idle.

Moreover, we set the cost of each instance by using the information provided in the Amazon EC2 Price History dataset\footnote{Amazon EC2 Spot Price History: \url{https://zenodo.org/doi/10.5281/zenodo.5880792}}.

We selected five Amazon EC2 instances among the all the available ones, namely \emph{d2.2xlarge}, \emph{m5ad.xlarge}, \emph{t2.medium}, \emph{t2.micro} listed in \Cref{tab:aws-instances}, as they represent conflicting trade-offs.
The selected instances have balanced characteristics in terms of performance, power consumption, and cost.
Thus, we avoid the algorithm to have a strong preference when searching the solution space.
In other terms, if an instance has the best value for each characteristic, the algorithm will always select it when searching the solution space because that instance will improve all the considered objectives.

\paragraph{Setting the algorithm} Once the source model has been built, we set up the optimization problem by defining objectives and constraints.
We set the objective function to minimize power consumption, response time, cost, and complexity.
We use the NSGA-II algorithm to build the Pareto front of the optimization problem.

We performed several trials to find the best configuration of the optimization problem. 
Finally, we set the size of the initial population of the NSGA-II algorithm to 16 individuals, the maximum number of generations to 200, and crossover and mutation probabilities to 0.8 and 0.2, respectively.
In our problem, the chromosome represents the sequence of refactoring actions that the designer can perform on the system.
We decided to limit this sequence length to four refactoring actions, because it seemed a reasonable number of actions that a designer can perform in sequence without leading the system architecture too far from the initial one. 

To mitigate the randomness of the optimization algorithm, we run the optimization process 31 times for each scenario, as suggested by \textcite{DBLP:conf/ssbse/ArcuriF11}.

\paragraph{Setting the experiments} In order to answer the research questions, we performed two different types of experiments, namely \base and \power.
The \base experiment does not consider power consumption as an objective of the optimization problem while keeping the other three (\ie response time, complexity, cost), whereas the \power one adds the power consumption to the objectives.

It is worth to note that, for sake of full comparison, the power consumption has been computed also for the solutions obtained in the \base, even though it has not been considered as an objective of the optimization problem.

 \section{What is the impact of sustainability on performance and cost? (RQ1)}
\label{sec:rq1}

By comparing a \base experiment with one that is \power, we aim at investigating how the objectives in our approach vary when power consumption is considered as an additional objective.
Clearly, we expect that, on average, the values of the power objective will be lower in the \power experiment because power is explicitly targeted by the optimization algorithm.
Nonetheless, it is important to understand how the power objective affects the other ones and, more concretely, how much we could trade in performance and cost for more sustainable solutions.
In this context, we still kept the \textit{complexity} objective to avoid considering disruptive sequences of refactoring actions on the initial architecture.

\subsection{Differences in the Pareto fronts}
\begin{figure*}[htpb]
    \centering
    \includegraphics[width=.98\textwidth]{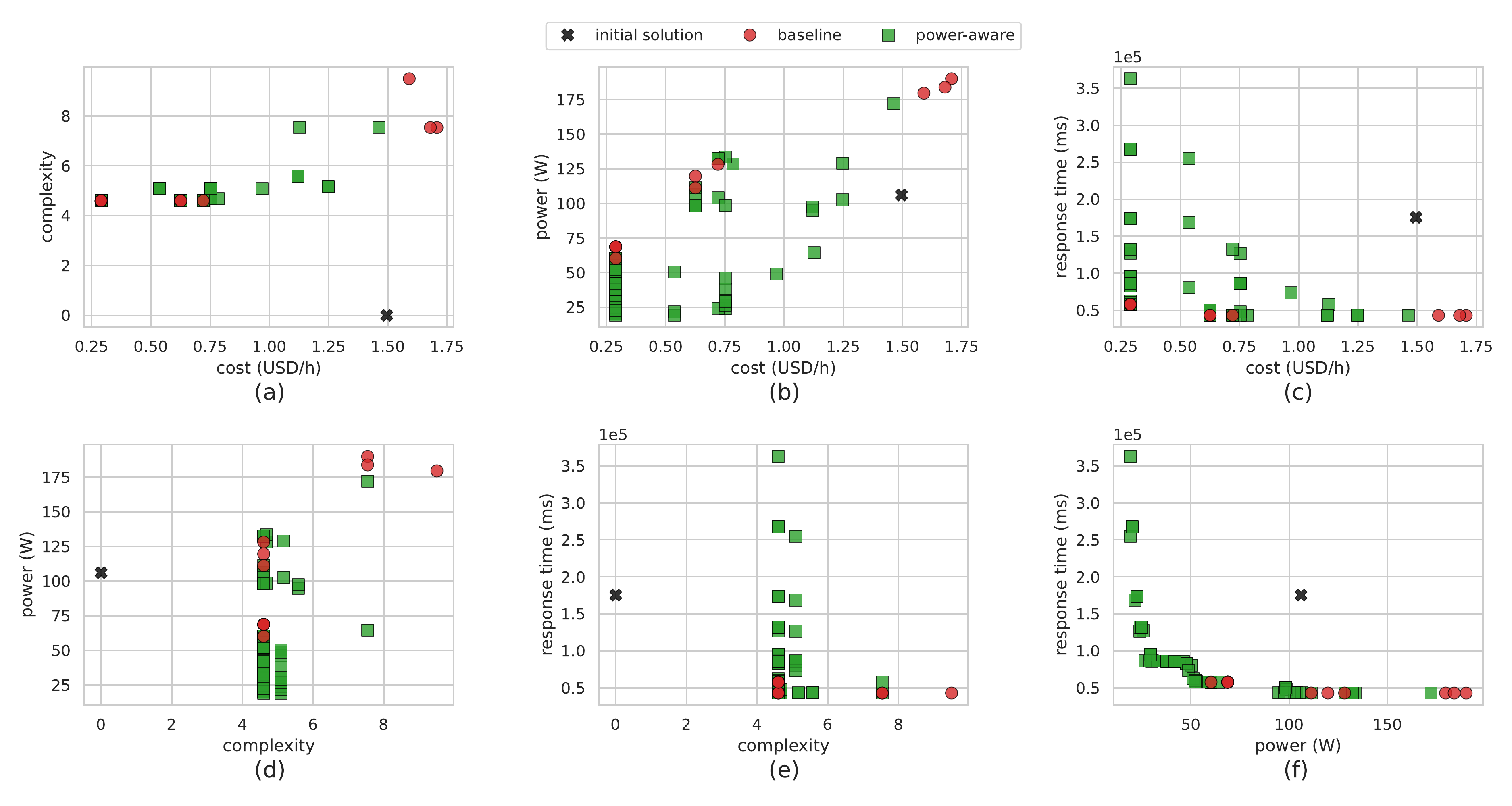}
    \caption{Comparison of the Pareto fronts resulting from the \base (without the power objective, 13 solutions) and the \power (with the power objective, 68 solutions) experiments. In the \base experiment, power consumption was not considered as an optimization objective, but only computed afterward on the models that are part of the Pareto front.}
    \label{fig:rq1_scatter}
\end{figure*}

\Cref{fig:rq1_scatter} shows the Pareto fronts obtained in the \base and \power experiments, which are obtained as the super-Pareto front of the 31 runs of each experiment. Namely, each solution in a super Pareto front is not dominated by any other solution in any of the 31 runs.

A first noticeable difference between the Pareto fronts is that the \power experiment has a larger number of solutions (68) than the \base experiment (13).
This is expected in most cases, because solutions found with a larger number of objectives live in a higher dimensional space where it is more difficult to be dominated.
A larger number of objectives is also likely to result in Pareto fronts which are more spread out in the space, as it is the case in our experiments.

The shapes of the Pareto fronts are quite noticeable for the \power experiment when \textit{power} and \textit{cost} are compared to \textit{response time} (Figures \ref{fig:rq1_scatter}c and \ref{fig:rq1_scatter}f, respectively), but not so much for the \base experiment.
Indeed, the \power experiment provides plenty of solutions to choose, with low power consumption, low response time, and a wide range of costs.
As expected, the \base is able to find solutions with low response time, but half of the Pareto front contains solutions with higher cost (Figure~\ref{fig:rq1_scatter}c).

Somehow unexpectedly, instead, both the experiments show a correlation between cost and power consumption (Figure~\ref{fig:rq1_scatter}b).
This is more evident in the \power experiment, and it is likely due to the trade-off in available cloud instances. Indeed, the current cloud offers, at a higher cost, instances with better performance and, most likely, higher power consumption.

Finally, the cost of changing the architecture (\textit{complexity}) does not vary so much across the two experiments (Figure~\ref{fig:rq1_scatter}a and \ref{fig:rq1_scatter}d).
By focusing on \textit{complexity}, Pareto fronts look more flat than in the other cases, most probably because this objective moves in discrete steps in the space, and not continuously as the other objectives. 

\subsection{Differences in the distributions of the objectives}

\begin{figure*}[htpb]
    \centering
    \includegraphics[width=\textwidth]{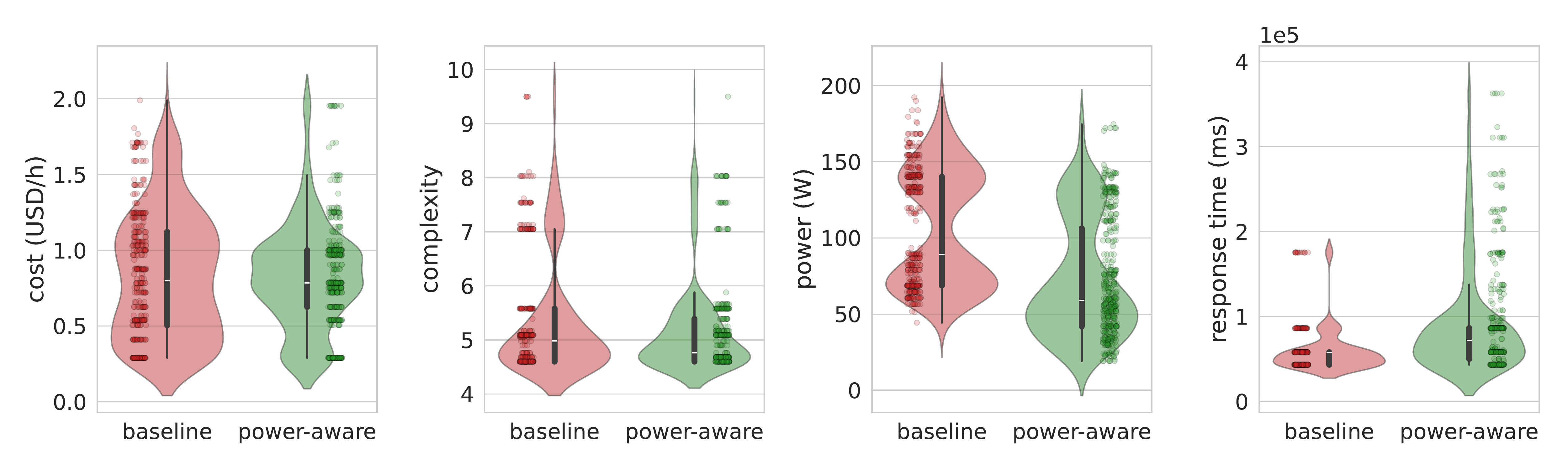}
    \caption{Distributions of the objective values in the \base and the \power experiments. Plotted solutions are all the Pareto fronts obtained in the 31 runs.}
    \label{fig:rq1_dist}
\end{figure*}

\Cref{fig:rq1_dist} shows the distributions of the values of the objectives in all the Pareto fronts of the 31 runs of the \base and \power experiments.
The distributions are shown as violin plots, with kernel density estimates, medians, and interquartile ranges as box plots inside violins.

While the \textit{cost} objective is significantly more spread out in the \base, the distributions of both experiments seem to be similar.
The \textit{complexity} objective exhibits a similar behavior in both experiments, with strikingly similar distributions.
Power consumption, while having similar distribution shapes, it is shifted in location towards lower values in the \power experiment, as expected.
Finally, the \textit{response time} objective is the one that shows the most significant difference across the two experiments.
In this case, the \power experiment has a higher median and a more spread out distribution, with a longer tail towards higher values, indicating a more diverse Pareto front with respect to this objective.

\subsection{Quantifying the effect of considering power consumption as an additional objective}
In order to provide a more precise estimation of the potential loss in performance and cost that may result from taking into account the additional objective of power consumption, we introduce the \textit{prospective sustainability penalty (PSP)} metric. 
Given an initial architecture and an optimization objective, PSP is defined through two characteristics: difference and magnitude.

\paragraph*{PSP difference}
This is an estimate of the difference in the objective values when the power objective is added to the optimization problem.
It is obtained by computing the Hodges-Lehmann estimator~\cite{hodges2011estimates} of the difference between the values of the objective in the \base and \power experiments.
The Hodges-Lehmann estimator is a robust estimator of the median of the distribution of the differences between two samples.
It is defined as the median of all the pairwise differences between the samples.
It is robust to outliers and does not require the distributions of the samples to be normal.
The formula for the Hodges-Lehmann estimator is:
    \begin{equation}
        \text{HL}(X, Y) = \text{median}\left(\left\{x_i - y_j \mid x_i \in X, y_j \in Y\right\}\right)
    \end{equation}
    In our case, $X$ and $Y$ are the values of the objective in the Pareto fronts of the \base and \power experiments, respectively.
    
\begin{table*}[htpb]
    \centering
    \caption{Descriptive statistics and prospective sustainability penalty (PSP) of the objectives in the \base and \power experiments. PSP is computed as the Hodges-Lehmann estimator of the difference between the values of the objective in all the Pareto fronts of the 31 runs of the \base and \power experiments, and the Cliff's delta.}
    \label{tab:psp}
    \begin{tabular}{lrrrrl}                     
\toprule
objective & mean difference & HL & MWU p-value & Cliff's delta & PSP \\
\midrule
cost (USD/h) & -0.185940 & 0.000000 & 0.737400 & -0.013435 & 0.00, -0.01 (negligible) \\
complexity & -0.621584 & 0.000000 & 0.546571 & 0.023641 & 0.00, 0.02 (negligible) \\
power (W) & -49.568407 & -30.318120 & 0.000000 & -0.529768 & -30.32, -0.53 (large) \\
response time (ms) & 38611.479219 & 14684.824000 & 0.000000 & 0.459672 & 14684.82, 0.46 (medium) \\
\bottomrule
\end{tabular}
 \end{table*}
\smallskip
\paragraph*{PSP magnitude}
This is an estimate of the magnitude of the differences in the objective values when the power objective is added to the optimization problem.
This is obtained by first performing the Mann-Whitney U test~\cite{mann1947test} on the values of the objective obtained for the Pareto fronts of each run in the \base and \power experiments, and then by computing a measure of effect size.
The Mann-Whitney U test is a non-parametric test that can be used to determine whether two independent samples were drawn from a population with the same distribution.
The null hypothesis of the test is that the two samples were drawn from the same distribution.
The alternative hypothesis is that the two samples were drawn from distributions with different medians.
The test returns a p-value, which is the probability of observing the given samples if the null hypothesis is true.
As it is customary, we set the threshold of the p-value to $0.05$.
If the null hypothesis is rejected, thus meaning that there was a significant difference between the \base and \power experiments, then the PSP magnitude is defined as the Cliff's delta~\cite{cliff1993dominance}.
The Cliff's delta is a non-parametric effect size measure that can be used to quantify the magnitude of the difference between two groups.
It ranges from -1 to 1, with positive values indicating a tendency for the first group to have larger values, and negative values indicating a tendency for the second group to have larger values.
The formula to compute the Cliff's delta ($\delta$) from the Mann-Whitney U statistic ($U$) is:
\begin{equation}
   \delta = \frac{2U}{n_1n_2} - 1
\end{equation}
where $n_1$ and $n_2$ are the sizes of the two samples. If the null hypothesis is not rejected, the PSP magnitude is defined as $0$.
We interpret the Cliff's delta as follows: $\delta < 0.147$ is a negligible effect, $0.147 \leq \delta < 0.33$ is a small effect, $0.33 \leq \delta < 0.474$ is a medium effect, and $\delta \geq 0.474$ is a large effect~\cite{romano2006appropriate}.

\smallskip
Therefore, given an initial architecture $A_0$ and the additional optimization objective $power$, the PSP of the objective $obj$ is defined as a pair:
\begin{equation}
    \text{PSP}_{power}(A_0, \text{obj}) = [\text{HL}(X, Y), \delta]
\end{equation}
where $X$ and $Y$ are the values of the objective in the Pareto fronts of the experiment with and without $power$, respectively.
\Cref{tab:psp} reports the $\text{PSP}_{power}$ for each objective, along with the difference in the mean, and the other statistics that were used to compute the PSP.
The \textit{power} objective obviously has a large PSP, as expected.
Other than that, the \textit{response time} objective has a medium PSP, with a Cliff's delta of $0.46$, which is very close to the threshold of $0.474$ that separates medium and large effects.

\medskip
\paragraph*{Summary}
In our experiments on the Train Ticket Booking Service case study, we observed that the addition of the power objective to the optimization problem has a significant impact on the response time of the system.
However, contrary to what one might expect, the impact on the deployment cost is negligible.
This leads to assume that, in order to obtain more sustainable solutions, we would have to trade in performance, but not in cost.
 \section{What is the effect of refactoring actions on the distribution of power and cost across different types of requests? (RQ2)}
\label{sec:rq2}

Individual functionalities contribute to the overall power consumption and cost of a system in different ways, because they are used with different frequencies and intensities, and they need different resources to be executed.
Each functionality can be associated to a type of request (\ie a scenario or a Sequence Diagram in UML), and the power consumption and cost of a type of request can be computed by aggregating the utilization of the resources that are used to serve that type of request.
In this section, we describe how we attribute power consumption and cost to individual types of requests.
Such information can be used to study how these two properties relate to user behavior, and provide an additional view on the trade-offs that should be considered when optimizing the system.

\subsection{Attributing power consumption and cost to types of requests}
We associate power consumption and cost to individual types of requests by adapting the models used for the optimization in \Cref{sec:approach:objective_computation}.
The amount of time that an instance is busy serving a type of request can be used to derive the share of the system power and cost that is spent throughout the system to serve that request.
This information can be obtained by reconstructing the flow of requests in the LQN that are generated from our UML architectures.
\begin{figure}[htpb]
    \centering
    \includegraphics[width=\linewidth]{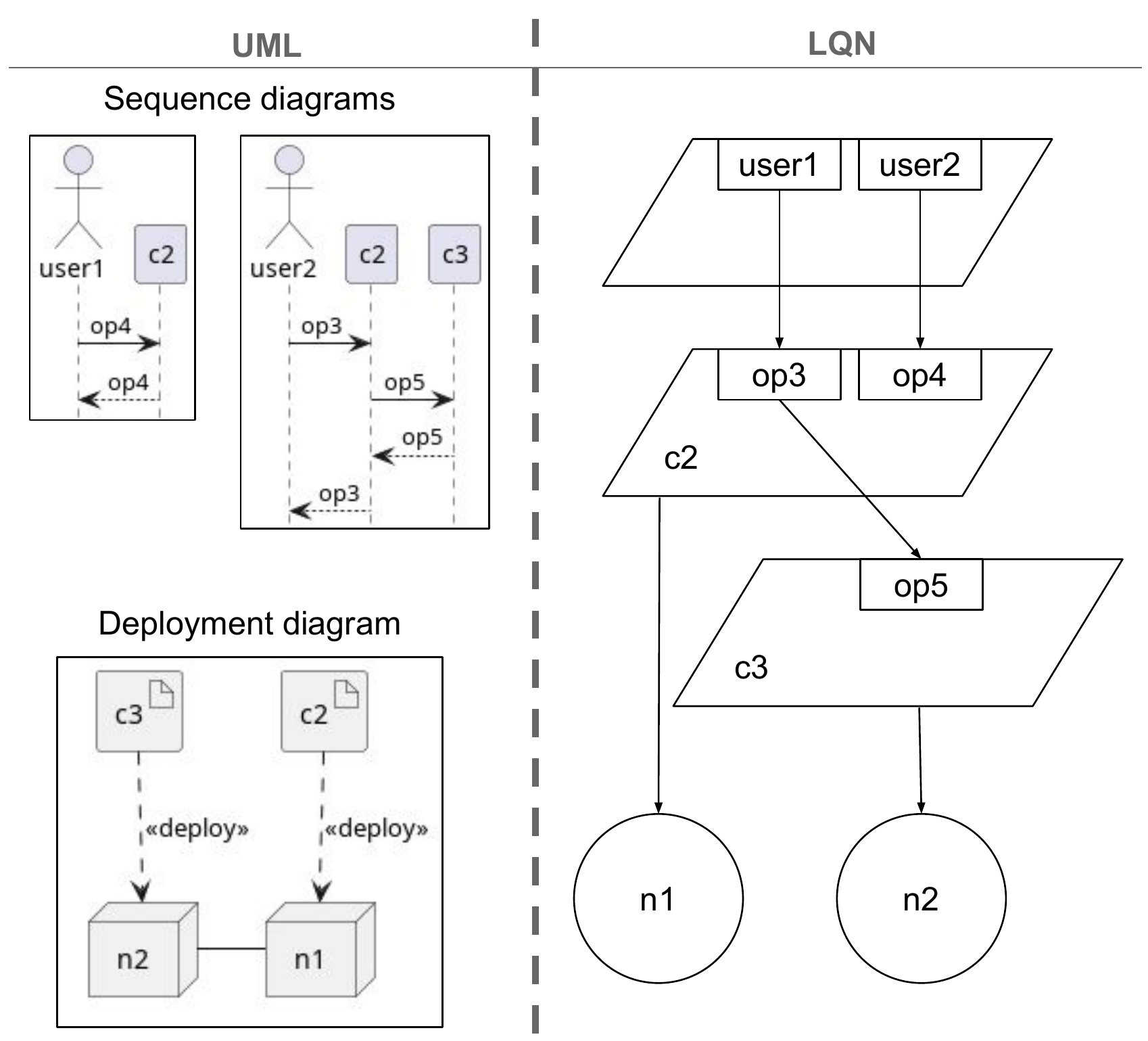}
    \caption{Simplified view of the transformation from UML to LQN.}
    \label{fig:uml2lqn}
\end{figure}
In the specification of our architecture, different types of requests are modelled through UML Sequence and Deployment Diagrams, and UML Nodes represent cloud instances.
As schematized in \Cref{fig:uml2lqn}, a UML Node is translated into a \textit{processor} in LQN, a UML Component into a \textit{task}, and a message in a Sequence Diagram becomes an \textit{entry} in the task.
When the LQN model is solved to compute performance indices, the solver annotates such indices back on the model.
Among the indices that the solver can compute, we are interested to the utilization of processors ($U_p$), and in particular to the share of this utilization associated to individual entries deployed on the processor ($U_{e,p}$).
Given a processor $p$ and an entry $e$, the contribution of an entry to the power consumption of the processor it is deployed on can be defined as:
\begin{multline}
    \text{power}(e) = U_{e,p} * \text{power}_{max}(p) + (1 - U_p) \\ \cdot k \cdot \text{power}_{max}(p)) \cdot U_{e,p} / U_p
\end{multline}
where $\text{power}_{max}(p)$ is the power consumption of the processor $p$ when busy, and $k$ is the factor by which scaling the power when the processor is idle.
The idea is to use the utilization of an entry to compute its power consumption when busy, and then sharing the idle time of the processor proportionally among the entries that are using it~\cite{Xu2023}.
The flow of a type of requests $r$ can be defined as a sequence $S=<e_1, e_2, ..., e_n>$ of entries $e$, each deployed on a processor $p$.
Accordingly, the power consumption of a type of requests $r$ can be defined as:
\begin{equation}
    \text{power}(r) = \sum_{i=1}^{|S|} \text{power}(e_i) \qquad \forall e_i \in S
\end{equation}
that is the sum of the power consumption of all the entries invoked to satisfy the request $r$ triggered by a given scenario.
\begin{figure*}[htpb]
    \centering
\includegraphics[width=\textwidth]{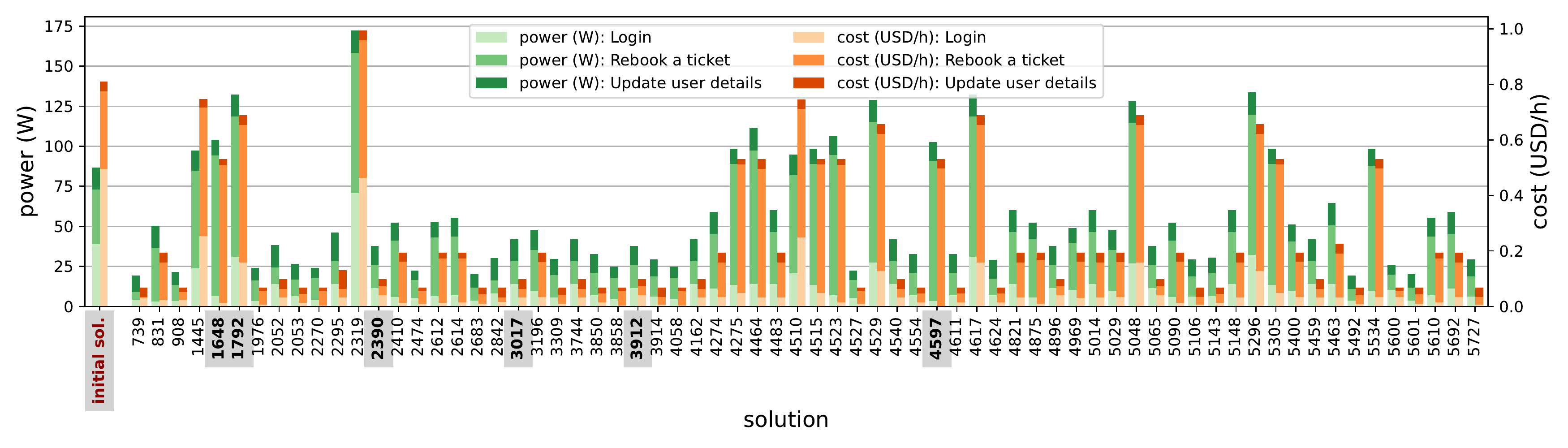}
    \caption{Power consumption and cost of individual types of requests for the solutions in the Pareto front of the \power experiment. Entire bars represent the total power consumption and cost of the system for a given solution. The x-axis lists solutions IDs, highlighted in bold if mentioned in the text.}
    \label{fig:power_and_cost}
\end{figure*}
A similar, but simpler, reasoning can be applied to compute the cost of a type of requests.
While the operating cost of the entire systems is computed as the sum of the cost of all the deployed nodes, the utilization of entries can be used again to attribute shares of that cost.
Therefore, the cost of a type of requests $r$ can be defined as:
\begin{equation}
    \text{cost}(r) = \sum_{i=1}^{|S|} \text{cost}(e_i) \qquad \forall e_i \in S
\end{equation}
where $\text{cost}(e_i) = \text{cost}(p) \cdot U_{e,p} / U_p$.

\subsection{User profiles}

\Cref{tab:rq2_power_and_cost} shows the mean, standard deviation, and median of power consumption and cost of individual types of requests (\textit{Scenarios}) in the super Pareto front of the \power experiment, along with the values for the initial solution.
On average, both cost and power consumption are reduced by the optimization, but with a high variability across the different types of requests.
The standard deviation of the \textit{Rebook a ticket} scenario is particularly high, most probably because this scenario is more complex than the other ones, and therefore more sensitive to changes in the architecture.
With such differences, it is important to let the designer aware of all the different trade-offs that can be made when optimizing the system, and that are not visible when looking at the overall power consumption and cost of the system.

\begin{table}[htpb]
    \caption{Summary of power consumption and cost of individual types of requests for the initial solution and for the solutions in the super Pareto front of the \power experiment.}
    \label{tab:rq2_power_and_cost}
    \begin{tabular}{lrrrr}
\toprule
\multirow{2}{*}{Scenario} &
\multirow{2}{*}{\thead{Initial\\solution}} &
\multicolumn{3}{c}{\power experiment} \\ \cmidrule{3-5}
&& mean & std & median \\
\midrule
cost: Login & 0.495 & 0.045 & 0.073 & 0.032 \\
cost: Rebook a ticket & 0.279 & 0.166 & 0.182 & 0.088 \\
cost: Update user details & 0.036 & 0.029 & 0.009 & 0.033 \\
power: Login & 38.791 & 11.610 & 10.049 & 9.727 \\
power: Rebook a ticket & 34.127 & 33.913 & 28.264 & 19.828 \\
power: Update user details & 13.815 & 11.441 & 2.508 & 11.883 \\
\bottomrule
\end{tabular} \end{table}

\Cref{fig:power_and_cost} shows a more detailed view of the same experiment.
It is important to remind that the optimization algorithm does not alter the number of requests of each type or their frequency, but only the way they are served by applying the refactoring actions defined in Section \ref{sec:approach:refactoring_catalog}.
Nonetheless, as it can be seen in the figure, the share of power consumption and cost of individual types of requests varies significantly across the solutions in the Pareto front.
This is due to the fact that the optimization impacts on the utilization of the resources that are used to serve each type of request and, consequently, on their shares of power consumption and cost.

In some configurations, a single type of request appears to be responsible for most of the power consumption and cost of the entire system, as it is the case for the \textit{Rebook a ticket} scenario in solutions with higher values of power and cost (\eg solutions 1648, 1792, 4597).
Conversely, there are cases with lower values of power and cost that are characterized by more balanced shares among the requests (\eg solutions 2390, 3017, 3912).

These cases in which different types of requests end up contributing to the overall power consumption and cost of the system in opposite ways would be very hard to spot without the support of the optimization, whereas they are fundamental to study the trade-offs that can be made when optimizing the system. For example, this awareness would allow architectural designers to decide that the cost and power consumption of a certain type of request is too high with respect to its priority/relevance in the system and, as a consequence, a limitation on the concurrent number of these requests can be introduced.

One possible reason for those disruptive changes in terms of power consumption and cost might be pinpointed to refactoring actions that saturate the utilization of some nodes in favor of cost reductions.
In other terms, the tendency of the optimization to minimize the number of nodes in order to reduce the cost of the system has the drawback of increasing their utilizations, thus possibly causing an increase in global power consumption.

\medskip
\paragraph*{Summary}
We have shown how the power consumption and cost of individual types of requests can be obtained on the basis of the utilization of the resources that serve them.
This information can be used to study how these two properties relate to user behavior, thus to provide an additional view on system trade-offs that are not visible when looking at the overall power consumption and cost of the system. Such a view enables architectural decisions aimed at fine-tuning the requirement satisfaction of different types of users.
 \section{How do the architectural solutions change when introducing power consumption among the optimization objectives? (RQ3)}\label{sec:rq3}

In this research question, we take a look at how the optimization algorithm employs the refactoring catalog at its disposal (\Cref{sec:approach:refactoring_catalog}) for sake of sustainability.
Indeed, we intend to extract possible insights about refactoring actions that seem to be beneficial in reducing power consumption.
Specifically, we focus on refactoring actions that are more frequently used by the algorithm and how they correlate with the objectives.

\begin{table}[htpb]
    \setlength{\tabcolsep}{4pt} \centering
    \caption{Comparison of the frequencies of types of refactoring actions in the super Pareto fronts of the \base and \power experiments. Target elements of the actions (\textit{Target} column), and where those elements were relocated (\textit{To} column) report the type of UML element as Node (N), Component (C), and Operation (O). Frequency column also shows the number of occurrences of the action type in the super Pareto front. Frequencies are reported for both experiments when the refactoring action and its target element are the same.}
    \label{tab:refact_freq_power}
    \begin{tabular}{lllll}
\toprule
\multirow{2}{*}{Type} & Target  & To    & \multicolumn{2}{c}{Frequency} \\
\cline{4-5}                                                                                                                               
                      & (N,C,O) & (N,C) & baseline & power-aware \\
\midrule
DROP & (N) verification & --- & 25.00\% (13) & 24.26\% (66) \\
DROP & (N) login & --- & 21.15\% (11) & 21.69\% (59) \\
DROP & (N) order-other & --- & 19.23\% (10) & 24.63\% (67) \\
DROP & (N) route-plan & --- & 13.46\% (7) & 15.81\% (43) \\
REDO & (C) order-other & (N) new-node & 7.69\% (4) & 0.74\% (2) \\
DROP & (N) travel-plan & --- & 3.85\% (2) & 2.57\% (7) \\
MOVE & (O) login & (C) ticket-info & 1.92\% (1) & 0.37\% (1) \\
MOVE & (O) updateuser & (C) travel-plan & 1.92\% (1) & --- \\
DROP & (N) rebook & --- & 1.92\% (1) & --- \\
DROP & (N) sso & --- & 1.92\% (1) & --- \\
DROP & (N) ticket-info & --- & 1.92\% (1) & 0.74\% (2) \\
MOVE & (O) login & (C) verification & --- & 1.47\% (4) \\
CLON & (N) login & --- & --- & 1.47\% (4) \\
MOVE & (O) getbyid & (C) rebook & --- & 1.47\% (4) \\
MOVE & (O) login & (C) rebook & --- & 1.10\% (3) \\
DROP & (N) seat & --- & --- & 1.10\% (3) \\
MOVE & (O) login & (C) travel-plan & --- & 0.74\% (2) \\
MOVE & (O) rebook & (C) order-other & --- & 0.74\% (2) \\
CLON & (N) ticket-info & --- & --- & 0.37\% (1) \\
MOVE & (O) login & (C) sso & --- & 0.37\% (1) \\
MOVE & (O) modify & (C) station & --- & 0.37\% (1) \\
\bottomrule
\end{tabular}
 \end{table}

\Cref{tab:refact_freq_power} shows the frequencies of the refactoring actions in the super Pareto fronts of the \base and \power experiments.
The removal of a node (\textit{DROP}) is by far the most frequent action in both experiments.
This is not surprising, as the removal of a node tends to reduce not only the power consumption of the system, but also its deployment cost.

When a node is removed, the approach has to relocate its components to other nodes, by preferring nodes hosting components that more frequently communicate with relocated ones. As explained in Section \ref{sec:approach:refactoring_catalog}, we have introduced such relocation criterion with the intent of not remotely spreading highly interacting components. 
The occurrence of so many removals of nodes might also indicate that the approach is able to identify nodes with a too low utilization to justify their cost.
For instance, the component responsible for the CAPTCHA verification is initially hosted on the \textit{verification} node. The latter is frequently removed and its component relocated to the \textit{login} node, which is the only node that communicates with the \textit{verification} node. The frequent occurrence of such refactoring action may indicate that the system is oversized and, eventually, the two components can be merged in one microservice. 

When we compare the frequencies of the action that redeploys a component to another node (\textit{REDO}), we can see that it is considerably more frequent in the \base experiment, and that the component is relocated to a new node, thus adding a new node to the architecture.
Conversely, when power consumption is considered, the approach tends to remove nodes instead of adding them, even if it may have a detrimental effect on the performance of the system.

The \textit{MOTN} action, which moves an operation to a new component on a new node, does not appear in the super Pareto front of either experiment.
This is probably due to the fact that the creation of a new node is a complex operation, and the approach tends to avoid it when possible.

Finally, we can see from the bottom half of \Cref{tab:refact_freq_power} that the \power experiment leads to a larger diversity in the number of refactoring actions, and to actions that are not present in the \base experiment.
From those actions that are specific to the Pareto front of \power experiment we can see, for instance, that the \textit{login} component and node are often  targets of different types of refactoring actions.
This confirms that the login functionality has room for improvement with regard to power consumption.

\paragraph*{Summary}
We have observed that the refactoring actions that most frequently occur in the Pareto fronts are the ones that remove nodes from the architecture.
This indicates that the algorithm is able to identify nodes that are hosting microservices that are too small to justify the cost and power consumption of a separate node.
We have also seen that, when the power consumption is considered, different types of refactoring actions and target elements are used, which leads to discover architectures that were not appearing in the \base experiment. \section{Related Work}\label{sec:related}

In the last decades, the problem of assessing sustainable systems is gaining more and more attention, as witnessed by the increasing number of studies in the literature~\cite{spe.2409,10.1145/2788397,10.1007/978-3-031-42592-9_4,HOUSSEIN2021100841}.

\textcite{spe.2409} highlighted challenges in designing systems, ranging from embedded systems to IoT devices.
Additionally, the study pointed to untapped research potential in green computing, energy-efficient systems, mobile cloud computing, and the Internet of Things within the context of architecting cloud-based systems. 
\textcite{10.1007/978-3-031-42592-9_4} addressed the crucial need to integrate sustainability aspects into architectural decision-making. 
Beyond technical expertise, architecture knowledge requires practical experience in representing, communicating, and managing architectural decisions.
Our approach exploits software architectures and gives to the designer a way to understanding the impact of sustainable constraints on software architectures.

\textcite{6976608} codified a Green Architectural Tactics catalog, providing architects with a systematic framework to incorporate energy-efficient design principles and deploy reusable solutions for developing environmentally conscious software.
\textcite{Ponsard2016398} introduced a UML profile aimed at augmenting the UML with energy-aware concepts.
We differentiate from the \citeauthor{6976608} study by introducing the automation concepts in applying refactoring actions on software architectures. 
Moreover, our refactoring actions could be mapped to some tactics in such catalog.
Instead of using the \citeauthor{Ponsard2016398} UML profile, we exploited the MARTE profile by the OMG group~\cite{MARTE}, which is considered the standard profile for non-functional analysis with UML.

\textcite{HOUSSEIN2021100841} presented a comprehensive review, categorizing popular meta-heuristic techniques based on scheduling nature, objectives, task-resource mapping, and constraints. 
Task scheduling is crucial for optimal cloud service performance, yet improper scheduling can lead to resource underutilization or overutilization, resulting in wastage or degraded service. To address these challenges, meta-heuristic algorithms have been incorporated into task scheduling, efficiently distributing diverse tasks across limited resources.
\textcite{10.1145/2788397} conducted a systematic mapping study to identify the state of the art of the optimization of virtualized resources within the cloud computing architecture. 
Moreover, \citeauthor{10.1145/2788397} established a taxonomy at two levels for scheduling cloud resources and systematically reviews state-of-the-art approaches.
\textcite{cpe.7112} introduced the Investment-Based Optimization (IBO) meta-heuristic algorithm to optimize scheduling while minimizing execution costs while maximizing load across computing resources.
Following these previous studies, we proposed a multi-objective approach to identify trade-offs between power consumption, response time, and cost while reducing refactoring actions complexity.
Furthermore, our approach considers software architectures as first class of citizen.

Besides, we proposed a metric (see \Cref{sec:rq1}) to evaluate the impact that sustainability constraints could have on other aspects, such as performance and cost.
The aim of such a metric is to provide to designers a quantifiable way to estimate improvement (detriment) that the sustainability has on their software systems. \section{Threats to validity}\label{sec:t2v}

\paragraph*{Construct validity}
We acknowledge that the parameters utilized for enhancing the initial models with non-functional information could have introduced variability into our results, as all objectives are influenced by such information.
To address this concern, we adopted model parametrization strategies from the literature~\cite{DBLP:conf/epew/WillneckerDBSKG15,DBLP:journals/jss/CortellessaPET22}.

Another potential source of impact on our study is the genetic algorithm employed in our experiments.
This algorithm plays a pivotal role in determining how solutions are generated by our approach, influencing the search space explored.
To minimize this potential threat, we opted for NSGA-II, a widely utilized algorithm~\cite{DBLP:conf/qosa/KoziolekKR11,DBLP:journals/smr/OuniKCSDI17} that proved to perform well in comparison with other multi-objective algorithms~\cite{DBLP:conf/cec/HiroyasuNM05}.

Additionally, we took precautionary measures aligned with established best practices in the literature~\cite{DBLP:conf/icse/ArcuriB11,DBLP:conf/ssbse/ArcuriF11} to alleviate the algorithm's influence on the final results.

\paragraph*{Internal validity}
Our exploratory study was designed as a paired comparison between an optimization experiment that does not consider power consumption and one that does.
A threat to the internal validity of our study is the possibility that the results obtained in RQ1 are not due to the introduction of power consumption as an optimization objective, but rather to the introduction of a new objective in general~\cite{Lai1994TOPSIS}.
To mitigate this threat, in RQ3 we contextualized the refactoring actions that were selected by the optimization algorithm, and we found that the actions selected in the \power experiment are more aligned with the goal of reducing power consumption than the actions selected in the \base experiment.

\paragraph*{External validity}
We only examined the results of the application of our approach to a single case study, albeit a large and widely used one.
Our intention was to provide a proof of concept of the applicability of our approach to a real-world system, and showcase how it can be used to support decision-making in the context of sustainable deployment planning.
However, we acknowledge that the results obtained may not be generalizable to other systems, and we plan further studies to validate the approach in other contexts.

Although our catalog of refactoring actions is relatively small, comprising only five actions, it is important to note that the complexity of potential modifications to the initial architectural model is not solely determined by the number of actions.
The challenge lies in the vast solution spaces created by the multitude of possible sequences in which these refactoring actions can be applied.
While a larger set of refactoring actions could theoretically capture more intricate changes, the associated increase in solution space would demand substantial computational resources, making it a trade-off between comprehensiveness and practical feasibility~\cite{DBLP:conf/cec/MartiGBM09}.

\paragraph*{Conclusion validity}
To mitigate the risk of drawing incorrect conclusions from our results, we adopted proper significance tests and effect size measures.
Specifically, we employed the Hodges-Lehmann estimator~\cite{hodges2011estimates} and the Cliff's delta effect size~\cite{cliff1993dominance} to avoid making assumptions about the distribution of the data.
We also reported p-values and effect sizes for all comparisons, and we only considered results with a confidence level of 95\%.
 \section{Conclusion}\label{sec:conclusion}

This paper introduces a novel approach that leverages NSGA-II to generate diversified deployment configurations of a software architecture through refactoring actions, while aiming to provide optimal trade-offs between system performance, deployment costs, complexity and power consumption.
The exploration of the intricate relationships between power consumption, cost, and user behavior is also supported, by looking at different types of user requests.

Results from our experiments on Train Ticket Booking Service application indicate that sustainability objectives, particularly focusing on power consumption, significantly impact system response time.
Surprisingly, this impact is observed with negligible effects on deployment costs.

As future work, we plan to evaluate the proposed approach on other case studies, and to investigate the impact of the proposed approach on the evolution of a software architecture over time.
Moreover, the awareness of the impact of deployment decisions on the cost and power consumption of serving certain types of requests opens to the possibility of using this information in a more fine-grained refactoring process that aims at optimizing towards specific user interaction scenarios.
Finally, by highlighting the variation in the contribution of individual types of requests, our approach could also identify opportunities for merging smaller microservices to save on power consumption and costs. 
\section*{Acknowledgments}
This work has been funded by \SoBigDataITAck, and \rechargeAck.

\printbibliography

\end{document}